%% file: main.tex
\documentclass[lettersize,journal]{IEEEtran}
\usepackage{amsmath,amsfonts}
\usepackage{algorithmic}
\usepackage{algorithm}
\usepackage{array}
\usepackage[caption=false,font=normalsize,labelfont=sf,textfont=sf]{subfig}
\usepackage{textcomp}
\usepackage{url}
\usepackage{stfloats}
\usepackage{verbatim}
\usepackage{multirow}
\usepackage{xcolor}
\usepackage{etoolbox}
\usepackage{graphicx}
\usepackage{acronym}
\usepackage{cite}
\usepackage{threeparttable}
\usepackage{orcidlink}
\input{Acronyms}

\usepackage{lipsum}
\usepackage{fancyhdr}

\makeatletter
\patchcmd{\@maketitle}
  {\addvspace{0.5\baselineskip}\egroup}
  {\addvspace{-1\baselineskip}\egroup}
  {}
  {}
\makeatother

\begin{document}


\title{Neural-HAR: A Dimension-Gated CNN Accelerator for Real-Time Radar Human Activity Recognition}

\author{Yizhuo~Wu\orcidlink{0009-0009-5087-7349},
    Francesco~Fioranelli\orcidlink{0000-0001-8254-8093},
    Chang~Gao*\orcidlink{0000-0002-3284-4078} \\
\thanks{*Corresponding author: Chang Gao (chang.gao@tudelft.nl)}
\IEEEauthorblockA{Department of Microelectronics, Delft University of Technology, Delft, The Netherlands}
}



\maketitle

\begin{abstract}
Radar-based human activity recognition (HAR) is attractive for unobtrusive and privacy-preserving monitoring, yet many \ac{CNN}/\ac{RNN} solutions remain too heavy for edge deployment, and even lightweight \ac{ViT}/\ac{SSM} variants often exceed practical compute and memory budgets. We introduce \textit{Neural-HAR}, a dimension-gated CNN accelerator tailored for real-time radar HAR on resource-constrained platforms. At its core is \textit{GateCNN}, a parameter-efficient Doppler–temporal network that (i) embeds Doppler vectors to emphasize frequency evolution over time and (ii) applies dual-path gated convolutions that modulate Doppler-aware content features with temporal gates, complemented by a residual path for stable training. On the University of Glasgow \texttt{UoG2020} continuous radar dataset, GateCNN attains $86.4\%$ accuracy with only $2.7$\,k parameters and $0.28$\,M FLOPs per inference, comparable to CNN–BiGRU at a fraction of the complexity. Our FPGA prototype on Xilinx Zynq-7000 Z-7007S reaches $107.5\,\mu$s latency and $15$\,mW dynamic power using LUT-based ROM and distributed RAM only (zero DSP/BRAM), demonstrating real-time, energy-efficient edge inference. Code and HLS conversion scripts are available at \url{https://github.com/lab-emi/AIRHAR}.
\end{abstract}


\section{Introduction}
\label{seq:Intro}

\IEEEPARstart{H}{uman} Activity Recognition (HAR) technologies have become critical in healthcare monitoring, elderly care, smart homes, and security applications~\cite{Survey-Continuous}. Among various sensing modalities, radar-based HAR offers a compelling alternative to wearable sensors and camera systems by preserving user privacy and comfort thanks to its contactless monitoring capability. The nature of radar sensing and its ability to sense also in non-line-of-sight conditions make it particularly attractive for continuous monitoring in ambient assisted living environments. However, deploying HAR models on resource-constrained edge devices and Field-Programmable Gate Arrays (FPGAs) remains challenging due to computational complexity, memory footprint, and power consumption constraints.

\begin{figure}
    \centering
    \includegraphics[width=\linewidth]{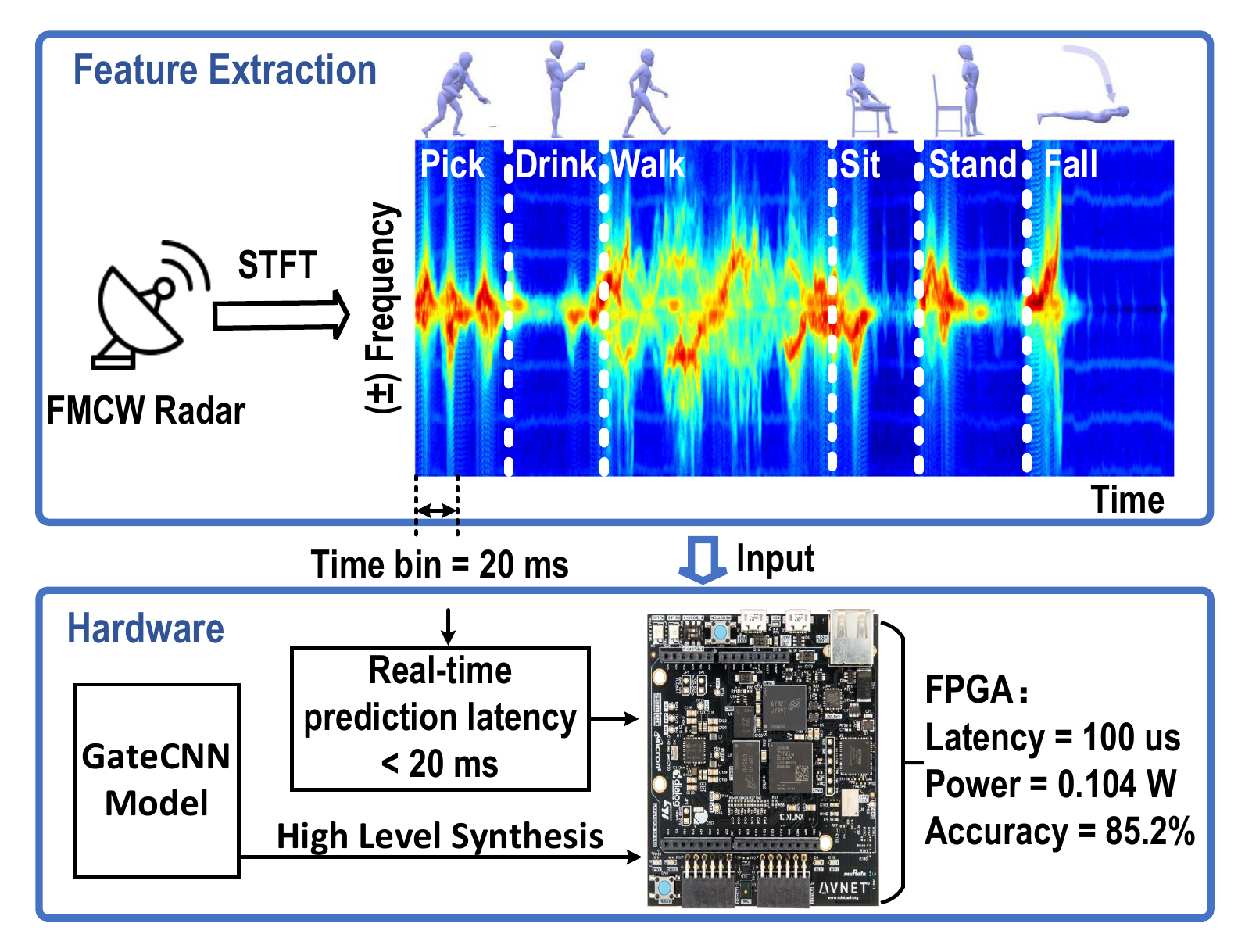}
    \caption{Hardware-oriented efficient GateCNN model for radar-based human activity recognition (HAR).}
    \label{fig:AIRHAR}
\end{figure}

Radar-based HAR primarily relies on micro-Doppler signatures, as shown in Fig.~\ref{fig:AIRHAR}, representing the time-frequency characteristics of radar echoes reflected from moving human body parts. These signatures capture the Doppler frequency shifts caused by the complex motion of different body segments during human activities, creating distinctive patterns that serve as unique fingerprints for activity classification. Micro-Doppler signatures encode rich information about human motion dynamics, including limb velocities, gait patterns, and temporal sequences of movement, making them highly discriminative features for distinguishing between different human activities.

Deep learning has substantially advanced radar-based HAR through automatic feature extraction from micro-Doppler signatures. Various neural network architectures have been explored, including CNN-based methods~\cite{Kim2016,CNN_TGRS,ComplexNN_TGRS,Yu2022}, RNN-based approaches~\cite{Bi-RNN,multimodal2020}, and hybrid CNN-RNN architectures~\cite{CNN-BiLSTM,CNN-LSTM,CNN-RNN-distributed}. While previous models achieve high classification accuracy across diverse activity recognition tasks, their computational complexity remains challenging for resource-limited applications such as mobile gesture recognition~\cite{linardakis2025} and distributed smart-home monitoring~\cite{DatasetD}.

Hardware implementation of deep learning models for radar-based HAR on FPGAs presents significant challenges at multiple levels. At the architectural level, recurrent architectures such as GRU and LSTM~\cite{GRU,LSTM} introduce sequential dependencies that prevent parallel processing and limit throughput. Hybrid CNN-RNN models, while achieving competitive accuracy through hierarchical feature extraction, require substantial resources with parameter counts around 71k~\cite{CNN-RNN-distributed} and arithmetic intensity exceeding 1G FLOPs per inference. These resource demands conflict with the constraints of edge deployment scenarios where power budgets and physical footprint are critical considerations. 

In this work, We present \textit{Neural-HAR}, a dimension-gated CNN accelerator for real-time radar HAR. Its backbone, \textit{GateCNN}, is designed around two observations: (1) micro-Doppler signatures contain complementary information along Doppler/velocity and time axes; (2) explicit modeling of Doppler evolution over time enables shallower networks with strong separability. GateCNN therefore (a) performs Doppler vector embedding to emphasize frequency change along time, and (b) uses dual-path gated convolutions in which a temporal path learns a gate that modulates Doppler-aware content features, with a residual connection preserving gradient flow. The resulting network is shallow and highly parameter-efficient, which translates to compact on-chip storage and simple datapaths.

We evaluate on the \texttt{UoG2020} continuous activity dataset~\cite{DatasetC} and prototype on a Xilinx Zynq Z-7007S. As shown in Fig.~\ref{fig:AIRHAR}, our contributions are:
\begin{itemize}
  \item Dimension-gated Doppler–temporal CNN. A lightweight architecture that attains $86.4\%$ accuracy with only $2.7$\,k parameters and $0.28$\,M FLOPs per inference, competitive with CNN–BiGRU while being markedly smaller and simpler.
  \item Real-time edge accelerator. An HLS-based \ac{FPGA} implementation achieving $107.5\,\mu$s latency and $15$\,mW dynamic power at 100\,MHz, storing all parameters in LUT-based ROM / distributed RAM with zero DSP and BRAM usage, validating practical, energy-efficient deployment.
\end{itemize}

Compared to prior CNN–RNN pipelines, Neural-HAR eliminates recurrent bottlenecks, enabling parallel-friendly hardware and deterministic low latency for continuous radar HAR at the edge.

\section{Proposed GateCNN}
\label{sec:gatecnn}

\begin{figure*}[!t]
    \centering
    \includegraphics[width=1.0\linewidth]{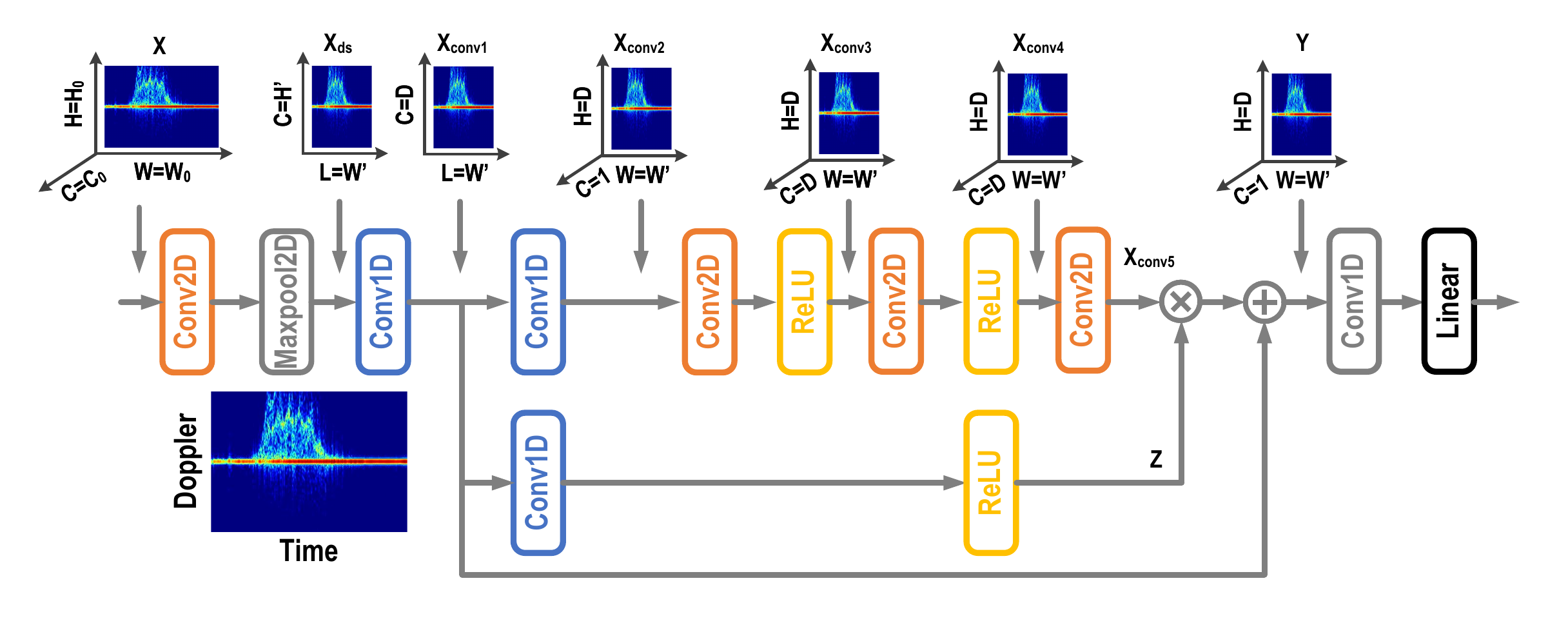}
    \caption{
    Architecture of the proposed GateCNN to process radar micro-Doppler signatures, i.e., 2D time-frequency maps of the target’s radial velocity over time (Doppler/velocity vs. time). The temporal path produces a gate, while the Doppler path extracts content features; the gate modulates Doppler-aware features, and a residual keeps gradient flow.}
    \label{fig:gatecnn}
\end{figure*}
The design of GateCNN is motivated by the observation that micro-Doppler signatures contain complementary information along temporal and Doppler/velocity dimensions. Traditional deep networks process these dimensions uniformly through hierarchical convolutions, requiring substantial depth to capture cross-dimensional interactions. In contrast, GateCNN emphasizes changes in Doppler/velocity information along time by 1D-convolution to explicitly model temporal-Doppler relationships, enabling efficient feature extraction with minimal parameters. 
As shown in Fig.~\ref{fig:gatecnn}, the architecture consists of dual-path gated projections that process features along orthogonal axes. In Fig.~\ref{fig:gatecnn}, $C,H,W$ represent the input channel, height, and width of the 2D convolution layer; $C,L$ represent the input channel and sequence length of the 1D convolution layers, respectively.

Given input micro-Doppler signatures $\mathbf{X}\in \mathbb{R}^{C_0 \times H_0\times W_0}$ where $C_0$, $H_0$, and $W_0$ represent channel, Doppler, and time dimensions respectively, the network first applies channel fusion and spatial downsampling to reduce the input dimensionality while preserving essential discriminative information:
\begin{align}
\mathbf{X}_1 &= \mathbf{W}_{c0} * \mathbf{X} \\
\mathbf{X}_{ds} &= \textit{MaxPool}(\mathbf{X}_1)
\end{align}
where $*$ denotes the convolution operation, $\mathbf{W}_{c0}$ represents learnable 2D convolutional kernels, and $\mathbf{X}_{ds} \in \mathbb{R}^{H' \times W'}$ denotes the downsampled feature map with $H' < H$ and $W' < W$. This initial stage reduces spatial dimensions while fusing channel information, establishing a compact feature representation suitable for subsequent processing.

Following the dimensionality reduction, Doppler-aligned convolution is applied along the Doppler axis to embed features in a learned representation space:
\begin{equation}
\mathbf{X}_{conv1} = \mathbf{W}_{c1} * \mathbf{X}_{ds}
\end{equation}
where $\mathbf{X}_{conv1} \in \mathbb{R}^{D \times W'}$ represents embedded features with output dimension $D$, and $\mathbf{W}_{c1}$ is an element-wise convolution kernel. Each channel spans the full Doppler dimension $H'$ at a single time instant, maintaining frequency continuity essential for human activity recognition. 

\subsection{Gated Convolutions}
\label{subsec:DSGP}
The core innovation of GateCNN lies right in its gating mechanism, which processes features through dual convolutional paths to capture temporal-Doppler interactions efficiently. Specifically, convolutions along the time axis first generate gate $\mathbf{Z}$ and content features $\mathbf{X}_{conv2}$ as:
\begin{align}
\mathbf{Z} &= \mathbf{W}_g * \mathbf{X}_{conv1} \\
\mathbf{X}_{conv2} &= \mathbf{W}_p * \mathbf{X}_{conv1}
\end{align}
where $\mathbf{W}_g$ and $\mathbf{W}_p$ are learnable kernels operating along the time dimension, and both $\mathbf{Z}, \mathbf{X}_{conv2} \in \mathbb{R}^{1 \times D \times W'}$. The gate features $\mathbf{Z}$ learn to identify salient temporal patterns, while the content features $\mathbf{X}_{conv2}$ undergo further processing to capture cross-dimensional relationships.

To explicitly model Doppler-domain patterns, the content features $\mathbf{X}_{conv2}$ are reshaped to emphasize the Doppler frequency change along time by Doppler vector embedding, enabling convolutions along the Doppler axis:
\begin{align}
\mathbf{X}_{conv3} &= \text{ReLU}(\mathbf{W}_{c2} * \mathbf{X}_{conv2}) \\
\mathbf{X}_{conv4} &= \text{ReLU}(\mathbf{W}_{c3} * \mathbf{X}_{conv3}) \\
\mathbf{X}_{conv5} &= \mathbf{W}_{c4} * \mathbf{X}_{conv4}
\end{align}
where $\mathbf{W}_{c2}$, $\mathbf{W}_{c3}$, and $\mathbf{W}_{c4}$ are 2D convolutional kernels. These cascaded convolutions process features in the Doppler frequency domain, capturing patterns that complement the temporal features extracted in the first path.

The two processing paths are then combined through the gating mechanism, which enables selective modulation of Doppler-processed features based on learned temporal gates:
\begin{align}
\mathbf{Y} &= \mathbf{X}_{conv5} \odot \text{ReLU}(\mathbf{Z}) + \mathbf{X}_{conv1}
\end{align}
where $\odot$ denotes element-wise multiplication and $\mathbf{Y} \in \mathbb{R}^{D \times W'}$. This gate modulates Doppler-processed features while the residual connection preserves gradient flow. 

The classification head aggregates spatial features through a learned averaging convolution along the Doppler dimension:
\begin{align}
\mathbf{v} &= \mathbf{W}_{avg} * \mathbf{Y} \\
\mathbf{\hat{y}} &= \mathbf{W}_{cls}\mathbf{v} + \mathbf{b}_{cls}
\end{align}
where $\mathbf{W}_{avg}$ is initialized to uniform weights, $\mathbf{v}$ is the flattened feature vector, and $\mathbf{W}_{cls}$ produces logits $\mathbf{\hat{y}} \in \mathbb{R}^{N_{cls}}$ for $N_{cls}$ activity classes.

\subsection{High-Level Synthesis Design}
\begin{figure}
    \centering
    \includegraphics[width=\linewidth]{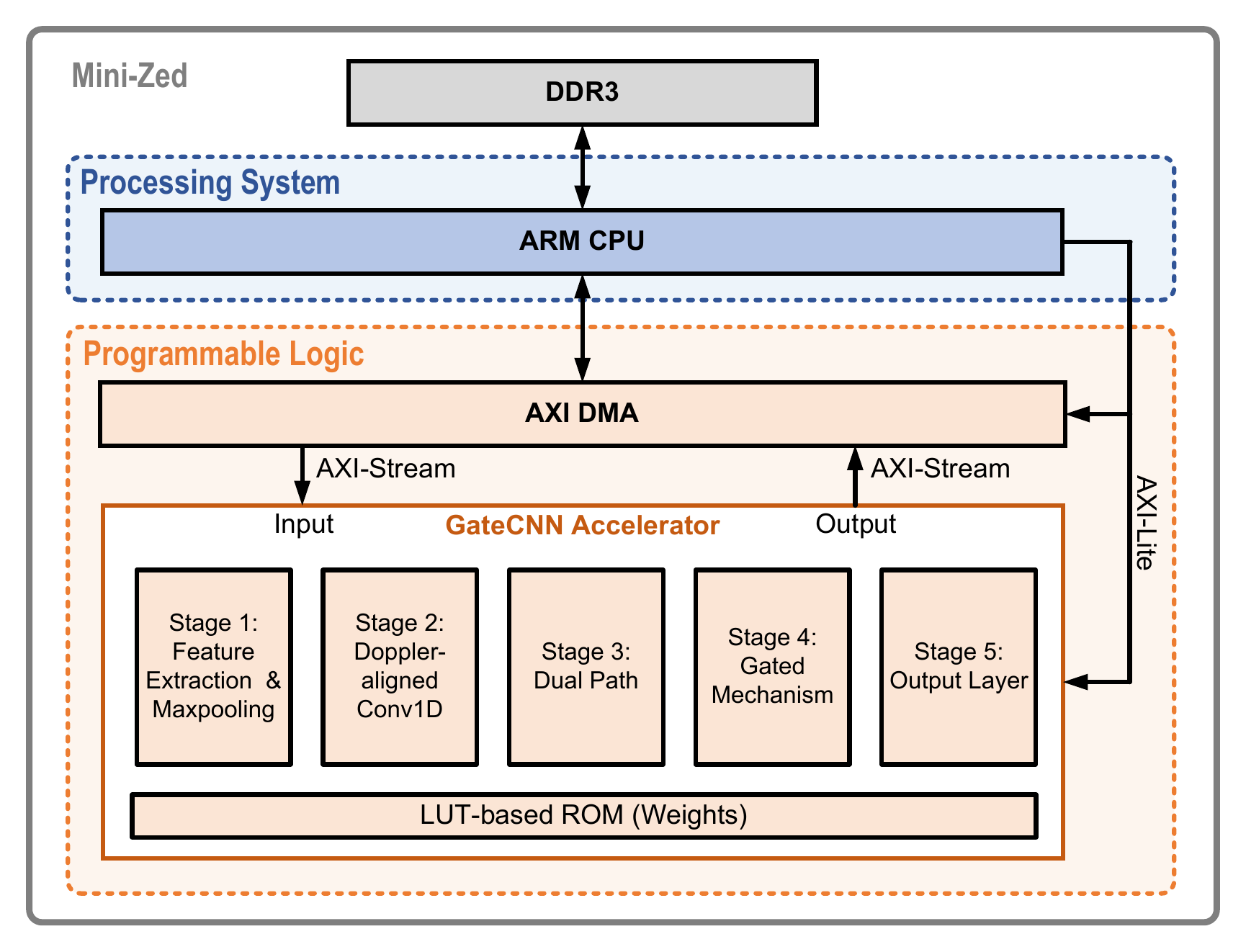}
    \caption{Proposed FPGA-based HAR System Design with HLS-based HAR Accelerator}
    \label{fig:hls_design}
\end{figure}

The GateCNN architecture was implemented using the hls4ml framework~\cite{fastml_hls4ml,Duarte:2018ite}, which provides automated translation from high-level neural network descriptions to optimized FPGA implementations. The conversion process begins with the trained PyTorch model exported to ONNX format, followed by automatic optimization including constant folding, shape inference, and channels-last format conversion. The hls4ml framework then generates optimized C++ code targeting Vitis HLS, with automatic precision quantization to 32-bit fixed-point arithmetic for efficient FPGA synthesis.

As shown in Fig.~\ref{fig:hls_design}, the HLS implementation employs a streaming architecture with dataflow pipeline processing, where the dual-path gating mechanism is realized through parallel processing paths: one generates gate features through temporal convolutions while the other processes content features through cascaded 2D convolutions. All network weights are stored as compile-time constants, enabling synthesis into LUT-based ROM without requiring external memory interfaces. The design operates at 100\,MHz clock frequency, achieving real-time processing capabilities suitable for continuous radar monitoring applications.

\section{Experimental Results}
\subsection{Dataset and Experimental Setup}
We evaluate GateCNN on the \texttt{UoG2020} continuous radar dataset~\cite{DatasetC}, where `continuous' refers to sequences of human activities performed consecutively without interruption. The dataset was acquired using a \ac{FMCW} radar operating at 5.8\,GHz with 400\,MHz bandwidth, comprising data from 15 participants (14 males, 1 female, aged 21–35) performing 6 activities within continuous 35-second sequences. The 6 activities include walking, sitting, standing, drinking, falling, and picking, representing typical scenarios in ambient assisted living applications. 
The micro-Doppler signatures are preprocessed into $(1,30,28)$ 2D frames (channel, Doppler bins, time steps) via short-time Fourier transform, with 2 participants held out for testing and the remaining 13 for training to ensure person-independent evaluation.

To validate practical deployment feasibility beyond algorithmic performance, GateCNN was implemented on a Xilinx Zynq-7000 Z-7007S FPGA, a resource-constrained device representative of edge computing platforms. The implementation follows a complete hardware design flow from high-level synthesis to post-place-and-route verification using hls4ml v1.1.0, Vitis HLS 2022.2, and Vivado 2022.2. The design operates at 100\,MHz clock frequency with 32-bit fixed-point precision. The latency is measured as the duration between the input valid and the output valid signals during behavioral simulation using one sample from \texttt{UoG2020} as test data.

\subsection{Comparison with Previous Works}

Table~\ref{tab:model_comp} presents a comparison between GateCNN and existing neural network architectures for radar-based HAR. All models were evaluated across 10 random seeds to ensure statistical significance.

\begin{table}[!t]
\caption{Mean and Standard Deviation of Classification Accuracy Performance across Seed 0 to 4 of Different NN-based RadHAR Models Evaluated With Dataset \texttt{UoG2020}~\cite{DatasetC} alongside Their Model Size and Floating-point Operations per Inference Sample ($\#$FLOP/Inf.)}
\label{tab:model_comp}
\resizebox{\linewidth}{!}{%
\begin{threeparttable}
\begin{tabular}{|ll|ccc|}
\hline \hline
\multicolumn{2}{|c|}{Classifiers} & \begin{tabular}[c]{@{}c@{}}\#params\\ (k)\end{tabular} & \begin{tabular}[c]{@{}c@{}}$\#$FLOP/Inf.~\tnote{a} \\ (M)\end{tabular} & \begin{tabular}[c]{@{}c@{}}Accuracy\\ (\%)\end{tabular} \\ \hline \hline
Bi-LSTM & ~\cite{DatasetC} & 3.0 & 0.034 & 85.6$\pm$1.42 \\
CNN-LSTM & ~\cite{CNN-LSTM} & 3.0 & 0.041 & 87.3$\pm$1.33 \\
CNN–BiGRU & ~\cite{CNN-RNN-distributed} & 3.1 & 0.711 & 88.4$\pm$1.58 \\\hline\hline
{\color[HTML]{9A0000} \textbf{GateCNN (Ours)}} & & {\color[HTML]{9A0000} \textbf{2.7}} & {\color[HTML]{9A0000} \textbf{0.28}} & {\color[HTML]{9A0000} \textbf{86.4$\pm$1.71}} \\\hline \hline
\end{tabular}%
\begin{tablenotes}
\item[a] Number of floating-point operations per inference (multiply–accumulate counted as two FLOPs).
\end{tablenotes}
\end{threeparttable}}
\end{table}

GateCNN with 2,719 parameters achieves 86.4\% accuracy, demonstrating competitive performance compared to existing architectures. The achieved accuracy is comparable to Bi-LSTM (85.6\%) while requiring fewer parameters, and approaches the performance of CNN-LSTM (87.3\%) and CNN–BiGRU (88.4\%) with a reduced number of parameters. Notably, GateCNN exhibits reasonable standard deviation ($\pm$1.71\%) across all evaluated models, indicating good training stability and robustness across different random initializations.

Beyond accuracy metrics, GateCNN demonstrates computational efficiency with 0.28\,M FLOPs per inference, achieved through efficient Doppler vector embedding and 1D convolution gated convolutions. While CNN-LSTM achieves higher accuracy (87.3\%) with lower FLOPs (0.041\,M), and CNN–BiGRU achieves the highest accuracy (88.4\%) with moderate FLOPs (0.711\,M), both hybrid CNN-RNN models suffer from sequential processing dependencies that fundamentally limit throughput. RNN-based architectures require sequential computation across time steps, preventing parallel processing and constraining maximum achievable throughput. 

\subsection{FPGA Implementation}

Table~\ref{tab:fpga_results} summarizes the implementation results across multiple metrics. Resource utilization is moderate, consuming 2,694 LUTs (18.71\%) and 2,694 registers (9.35\%) of the available resources on the Z-7007S device. Notably, the implementation requires zero DSP blocks and zero BRAM. The network parameters are stored as constants directly synthesized into LUT-based ROM, where the small parameter count (2,719 parameters $\times$ 32 bits $\approx$ 11\,KB) fits entirely within the distributed memory resources.

 \texttt{UoG2020} has 1750 time bins for each 35-second sequence. As each time bin of \texttt{UoG2020} is 20\,ms, the real-time prediction requires latency less than 20\,ms. The implemented FPGA latency is measured as 107.5$\mu s$, enabling real-time processing and leading to an achieved throughput of 9,302 inferences per second. This performance headroom provides opportunities to further minimize power consumption or maintain full throughput to support additional signal processing tasks, such as preprocessing. 

\begin{table}[!t]
\caption{FPGA Implementation Results on Xilinx Zynq Z-7007S}
\label{tab:fpga_results}
\centering
\begin{tabular}{|l|c|}
\hline
\textbf{Parameter} & \textbf{Value} \\
\hline \hline
Target Device & Xilinx Z-7007S \\
Clock Frequency & 100 MHz \\
Precision & 32-bit fixed-point \\
Inference Latency & 107.5 $\mu$s \\
Throughput & 9.3 kInf/s \\
\hline
LUT Utilization & 18.71\% (2,694) \\
FF Utilization & 9.35\% (2,694) \\
DSP Utilization & 0\% (0) \\
BRAM Utilization & 0\% (0) \\
\hline
Total Power & 0.104 W \\
Dynamic Power & 15 mW \\\hline
\end{tabular}
\end{table}

Power analysis reveals total on-chip power consumption of 0.104\,W, with dynamic power of only 15\,mW and static power of 90\,mW. This low dynamic power consumption is particularly attractive for battery-powered applications, as it represents the activity-dependent power overhead. The low power profile, combined with the small footprint, makes the implementation suitable for distributed radar sensor networks with strict energy budgets, such as smart home monitoring systems or wearable radar devices.

\section{Conclusion}
\label{sec:conclusion}
We presented \textit{Neural-HAR}, a dimension-gated CNN accelerator for real-time radar HAR. Its backbone, \textit{GateCNN}, couples Doppler vector embedding with dual-path gated convolutions to capture complementary temporal and frequency-domain cues using a compact, shallow network. On \texttt{UoG2020}, GateCNN delivers $86.4\%$ accuracy with only $2.7$\,k parameters and $0.28$\,M FLOPs per inference. The HLS-based prototype on Xilinx Zynq-7000 Z-7007S achieves $107.5\,\mu$s latency and $15$\,mW dynamic power without using DSPs or BRAM, demonstrating that accurate radar HAR can be performed on modest edge hardware with tight energy budgets. Future work will extend Neural-HAR to multi-radar fusion and event-driven streaming, and explore lower-precision quantization and on-chip learning for adaptive, long-term monitoring.

\bibliographystyle{IEEEtran}

\bibliography{IEEEabrv,ref.bib} 

\end{document}

%% file: Acronyms.tex
\newacro{HAR}{Human Activity Recognition}
\newacro{RadHAR}{Radar-based Human Activity Recognition}

\newacro{RF}{Radio Frequency}

\newacro{CW}{Continuous Wave}
\newacro{FMCW}{Frequency Modulated Continuous Wave}
\newacro{UWB}{Ultra-Wide Band}
\newacro{PRI}{Pulse Repetition Interval}

\newacro{STFT}{Short-Time Fourier Transform}
\newacro{SVD}{Singular Value Decomposition}
\newacro{ZOH}{Zero-Order Hold}

\newacro{DL}{Deep Learning}
\newacro{FLOP}{Floating-point Operation}
\newacro{ADAMW}{Adaptive Moment Estimation with decoupled Weight decay}

\newacro{1D}{One-Dimensional}
\newacro{2D}{Two-Dimensional}

\newacro{kNN}{k-Nearest Neighbors}
\newacro{SVM}{Support Vector Machine}

\newacro{CNN}{Convolutional Neural Network}
\newacro{RNN}{Recurrent Neural Network}
\newacro{Bi}{Bidirectional}
\newacro{LSTM}{Long Short-Term Memory}
\newacro{GRU}{Gated Recurrent Unit}

\newacro{Bi-LSTM}{Bidirectional Long Short-Term Memory}
\newacro{CNN-LSTM}{}

\newacro{ViT}{Vision Transformer}
\newacro{SSM}{State Space Model}
\newacro{ViM}{Vision Mamba}

\newacro{SiLU}{Sigmoid Linear Unit}
\newacro{FML}{Frequency Modulated Lightweight}
\newacro{LW}{Lightweight}

\newacro{HLS}{High-Level Synthesis}

\newacro{FPGA}{Field-Programmable Gate Array}

